# Using multiple representations to improve student understanding of quantum states

Emily Marshman[1], Alexandru Maries[2] and Chandralekha Singh[3]

[3] *Department of Physics & Natural Science, Community College of Allegheny County, Pittsburgh, PA, 15212, USA*
[3] *Department of Physics, University of Cincinnati, Cincinnati, OH, 45221, USA*
[3] *Department of Physics and Astronomy, University of Pittsburgh, Pittsburgh, PA, 15260, USA*

**Abstract.** One hallmark of expertise in physics is the ability to translate between different representations of knowledge and use the representations that make the problem-solving process easier. In quantum mechanics, students learn about several ways to represent quantum states, e.g., as state vectors in Dirac notation and as wavefunctions in position and momentum representation. Many advanced students in upper-level undergraduate and graduate quantum mechanics courses have difficulty translating state vectors in Dirac notation to wavefunctions in the position or momentum representation and vice versa. They also struggle when translating the wavefunction between the position and momentum representations. The research presented here describes the difficulties that students have with these issues and how research was used as a guide in the development, validation, and evaluation of a Quantum Interactive Learning Tutorial (QuILT) to help students develop a functional understanding of these concepts. The QuILT strives to help students with different representations of quantum states as state vectors in Dirac notation and as wavefunctions in position and momentum representation and with translating between these representations. We discuss the effectiveness of the QuILT from in-class implementation and evaluation.

## I. INTRODUCTION AND THEORETICAL FRAMEWORK

We are in midst of a second quantum revolution enabled by our ability to exquisitely control and manipulate quantum systems [1-9]. Quantum information science and technology is a rapidly growing interdisciplinary field with applications in quantum computing, communication, and sensing. There is an explosive growth in exciting career opportunities for tomorrow's quantum leaders. Helping physics students learn foundational concepts in this interdisciplinary field is vital for inspiring them to become future contributors in this exciting field. Quantum computers are powered by the most fundamental equation in quantum mechanics: the time-dependent Schrödinger equation. However, there are many competing physical quantum computing platforms, none of which have really crossed a threshold or have "broken from the pack" with clear ability to scale to truly useful sizes. Therefore, while for those interested, e.g., in quantum algorithms, a good grasp of the quantum formalism of a two-level system and multi-qubit systems is sufficient, those interested in hardware issues pertaining to making robust qubits for building and harnessing a scalable quantum computer must also consider practical issues involved in experimentation particularly because quantum states are fragile. The latter group must have a deep understanding of quantum physics beyond two-level systems. In particular, while many aspects surrounding quantum computing do not require any understanding of the physical underpinnings of these developing technologies, we are still far from being able to build fault tolerant quantum computers and fully abstract ourselves from the underlying quantum physics of designing robust qubits. Therefore, physics students interested in the second quantum revolution hardware issues and developing robust qubits that are fault tolerant must be guided through courses and curricula that are rich in quantum physics beyond the two-level systems.

Research shows that learning quantum mechanics is challenging for multiple reasons such as the fact that it involves abstract subject matter, a paradigm shift from classical mechanics and uses representations such as Dirac notation which can be unfamiliar even for advanced students. Several studies have focused on student difficulties after traditional lecture-based instruction [10-18] including quantum measurement [19-24], probability distributions for measuring observables, expectation values and their time dependence [25-28], addition of angular momentum [29, 30], and experiments involving quantum phenomena [31-34]. Reviews of common student difficulties and other issues focused on quantum mechanics [35, 36] have been published, while other studies have focused on facets of cognitive issues and reasoning difficulties and their relation to the paradigm shift from classical to quantum mechanics [37-41]. There is also research on other advanced concepts as well as bound and scattering states and black body radiation [42-50]. Researchers have discussed development of research-based instructional materials and pedagogies to help students learn quantum concepts better as in several earlier references cited here as well as in Ref. [51-59]. Researchers have also investigated other aspects of learning in the context of quantum mechanics, e.g., related to epistemological beliefs or confidence [60, 61]. Some investigations have focused on visualization [62-66], or how games can help students learn quantum mechanics [67]. Other studies have investigated attitudes of those who teach quantum mechanics [68, 69]. Research has also focused on developing conceptual surveys to investigate the extent to which students have developed a good grasp of different quantum concepts [70-75].

Furthermore, to develop expertise in quantum mechanics, students must develop a functional understanding of Dirac notation, and some prior studies have investigated student difficulties with Dirac notation [19, 76-82]. More broadly though,

Dirac notation is just one of the representations used to describe quantum concepts, and a functional understanding of quantum mechanics implies the ability to move flexibly between different representations, such as from a state vector written in Dirac notation (e.g., $|\Psi\rangle$) to the corresponding wavefunction in the position or momentum representation written in Dirac notation (e.g., as $\langle x|\Psi\rangle$, $\langle p|\Psi\rangle$ where $x$ and $p$ are the running indices), to the same wavefunctions written in algebraic notation (e.g., $\Psi(x)$, $\phi(p)$ written without Dirac notation) and vice versa. We note that for the sake of conciseness, throughout this paper, when we refer to "wavefunctions in position and momentum representation", we mean both when written using Dirac notation and algebraic notation as described above. In other words, as it has commonly been stressed in research on student learning in physics, the ability to express a concept in different representations is a hallmark of expertise [83-100]. This is because in physics (and science in general), "[t]here is no purely abstract understanding of a physical concept—it is always expressed in some form of representation."[83] This means that representations are part of the 'disciplinary discourse' in physics and being able to meaningfully participate in this discourse requires facility with the various representations of the physical concepts involved. As Airey and Linder put it [101, 102], there is a "critical constellation of semiotic resources that is needed for appropriate disciplinary knowledge construction", and they noted that, in physics, semiotic resources include "graphs, diagrams, sketches, figures, mathematics, specialist language, etc." So in order to learn physics effectively, students need practice with engaging in the disciplinary discourse which requires use of the representations most commonly used by experts to describe physical phenomena [101, 102]. In the language of social semiotics [103], translating between different representations of the same phenomena is referred to as transduction and it is precisely this (along with its close relative – transformation – the ability to relate similar representations of different concepts) that helps students develop an understanding of the affordances of the different representations. This type of facility requires that students have practice with using the various representations as well as translating between different representations of the same concept as well between similar representations of different concepts in order to progress towards expertise in quantum mechanics (and indeed, physics in general).

Throughout this paper, we use the term "representations" as defined by Meltzer, i.e., "the diverse forms in which physical concepts may be understood and communicated" [83]. Therefore, the term "representation" here refers to Dirac notation, as well as wavefunction in the position or momentum representation. This is very relevant for quantum mechanics, but very few prior research studies have focused specifically on issues related to representations in quantum mechanics [76, 104-106].

Gire and Price identified what they described as "structural features" of quantum representations (Dirac notation, matrix and algebraic notations) and discussed how student reasoning interacts with these features to guide their reasoning [104]. Based on student interviews, they found that the features of a representation guide students' approach to a problem using that representation. For example, the compactness, e.g., of states in Dirac notation supports reasoning for certain computations, and can be helpful for writing algebraic representations of quantum states [104]. Wawro et al. [76] discussed "metarepresentational competence" in quantum mechanics, which is defined as "the faculty to generate, critique, and refine representational forms." One aspect of metarepresentational competence is the ability to recognize the utility of one representation vs. another for performing a particular task; for example, Wawro et al. [76] suggest that Dirac notation in an energy basis is preferable when computing energy expectation values, but position representation is preferable when computing the probability that a particle in an infinite square well is found in one half of the well [76]. The authors also stressed that for students to be able to choose one representation vs. another, it's also important for them to be able to translate between different representations, which they viewed as another aspect of metarepresentational competence [76].

Here we expand upon an investigation of student difficulties with representing quantum state vectors in Dirac notation and as wavefunctions in the position or momentum representation and translating between these different representations [79]. We also describe how this research was used as a guide for the development, validation, and in-class evaluation of a research-based Quantum Interactive Learning Tutorial (QuILT) to help students learn these concepts, which includes learning to translate between these different representations. Other sections of the Dirac notation QuILT help students learn about operators [77], probability distribution [25] and expectation values [27]. Other QuILTs discussed previously include those involving the Stern-Gerlach experiment [31], quantum measurement [21], time-evolution [51], addition of angular momentum [30], the double-slit experiment [53], the Mach-Zehnder interferometer using single photons [32], basics of degenerate perturbation theory [42], intermediate field [48] as well as strong and weak field Zeeman effect [43], fine-structure of hydrogen atom [49], basics of identical particles [44], determining the number of distinct many-particle states, possible outcomes of energy measurement and probability of obtaining a particular energy if we randomly measured the energy of one particle for identical particles with fixed total energy [107], quantum key distribution [52], Bloch sphere [108] and basics of quantum computing [9].

This research was guided by the aforementioned recognition of the importance of representations in learning quantum mechanics as well as the framework of Zone of Proximal Development (ZPD) attributed to Vygotsky [109] consistent with other QuILTs [25, 27, 77]. The ZPD is defined as the difference between what a learner can achieve on their own without support and what they can achieve under the guidance of an expert or in collaboration with peers. This framework emphasizes that for meaningful learning to occur, the activities students should engage in to learn must be within their ZPD, which itself is dynamic and grows as learners progress in their level of expertise. Thus, effective instruction must be commensurate with students' prior knowledge and build on their knowledge at a given time. Engaging with carefully designed instructional tasks such as the guided inquiry-based teaching learning sequences in the QuILT in collaboration with peers or with guidance from instructors [110], can stretch students' ZPD and help them develop expertise in quantum mechanics. For instruction to incorporate students' initial knowledge appropriately, one must investigate student difficulties with relevant concepts and use this research as a guide in developing and validating learning tools such as the QuILT we discuss here. To ensure that the learning activities students engage with while working on the QuILT were in their ZPD, our investigation identified common student difficulties and used them in the design of the QuILT that strives to help students learn these concepts (described in detail in the next section).

## II. METHODOLOGY

As noted, the Dirac Notation QuILT has multiple sections which focus on helping students learn about other important aspects of using Dirac notation in quantum mechanics and we have written about student learning of those topics elsewhere [25, 27, 77]. Due to this, the methodology described in this article is similar to the methodology of those other articles, but is most similar to the methodology in Ref. [25].

### A. Participants

The participants in this study were:
- Undergraduate students enrolled in a mandatory upper-level (junior and senior physics majors) quantum mechanics course at a large public research university in the US:
  - For the investigation of difficulties as well as the evaluation of the QuILT (see next section for Q1-Q7), four cohorts from four separate semesters at a large public research university participated totaling 85 students.
  - In addition, for the retention question given at the end of the course (see next section), in the QuILT group, 58 students participated (a subset of the 85 from the large public research university) and in the non-QuILT group, a total of 184 students from four other large research universities participated.
  - 23 additional students from the large public research university participated in one-on-one think-aloud interviews (described in subsection C. Methods used to identify student difficulties). These students all had traditional instruction in the relevant concepts before participating in the interviews. We note that these students were different from the ones who were part of the quantitative data related to Q1-Q7 (described in the next section).
- Graduate students: The graduate students were first-year (first semester) Ph.D. students at the same large public research university and were simultaneously enrolled in a grad QM course and a Teaching Assistant training course. In some years the pre-test, tutorial, and post-test were administered in the Teaching Assistant training course and other years they were administered in the grad QM course depending on the schedules of the instructors. In total, 97 graduate students participated.
- Three faculty members who are experts in these topics and two graduate students who conduct research in student understanding of quantum mechanics provided feedback and suggestions at various stages of the development and validation of the QuILT.

### B. Learning objectives

The learning objectives for the QuILT are as follows (we note that for this article, all the questions discussed are in one dimension). Students should be able to:
- express a quantum state vector in Dirac notation as a wavefunction in position representation (position space wavefunction) and a wavefunction in momentum representation (momentum space wavefunction) using both Dirac and algebraic notations.
- describe how to obtain the wavefunction in position representation starting with a generic quantum state vector $|\Psi\rangle$.
- describe how to obtain the wavefunction in momentum representation starting with a generic quantum state vector $|\Psi\rangle$.
- describe how to obtain the wavefunction in momentum representation from the wavefunction in position representation.
- write a momentum eigenstate with eigenvalue $p'$ in position representation using Dirac notation.
- write a momentum eigenstate with eigenvalue $p'$ in momentum representation using Dirac notation.
- express $\langle x|x'\rangle$ written in Dirac notation as a position eigenfunction in position representation using algebraic notation.

- express $\langle p|p'\rangle$ written in Dirac notation as a momentum eigenfunction in momentum representation using algebraic notation.
- express $\langle x|p'\rangle$ written in Dirac notation as a momentum eigenfunction in position representation using algebraic notation.
- express $\langle p|x'\rangle$ written in Dirac notation as a position eigenfunction in position representation using algebraic notation.

To identify the requisite knowledge related to quantum states in Dirac notation and as wavefunctions in position and momentum representations, we used a cognitive task analysis from an expert perspective (see Table I). A generic quantum state $|\Psi\rangle$ (here in Dirac notation) includes all the relevant information pertaining to the system. Representing $|\Psi\rangle$ as a wavefunction in position representation requires projecting $|\Psi\rangle$ along position eigenstates $|x\rangle$: the position space wavefunction is $\langle x|\Psi\rangle = \Psi(x)$ (here $x$ is a continuous index). Similarly, representing $|\Psi\rangle$ as a wavefunction in momentum representation requires projecting $|\Psi\rangle$ along momentum eigenstates $|p\rangle$: the momentum space wavefunction is $\langle p|\Psi\rangle = \Phi(p)$ (here, $p$ is a continuous index). Similarly, representing a position eigenstate $|x'\rangle$ with eigenvalue $x'$ and a momentum eigenstate $|p'\rangle$ with eigenvalue $p'$ in the position or momentum representation requires projecting them along position eigenstates $|x\rangle$ (position basis) or momentum eigenstates $|p\rangle$ (momentum basis). In particular, if we ignore normalization, a position eigenstate in the position representation is a Dirac delta function $\langle x|x'\rangle = \delta(x-x')$ and a position eigenstate $|x'\rangle$ in the momentum representation is represented as $\langle p|x'\rangle = e^{-ipx'/\hbar}$. To translate between position and momentum representations, one must perform a Fourier transformation (or inverse Fourier transformation). For example, the wavefunction in position representation can be written in the momentum representation via the Fourier transformation $\Phi(p) = \langle p|\Psi\rangle = \int_{-\infty}^{\infty} \langle p|x\rangle \langle x|\Psi\rangle dx = \int_{-\infty}^{\infty} e^{-ipx/\hbar} \Psi(x) dx$.

**Table I**. Different quantum states as state vectors in Dirac notation, as wavefunctions in position representation and momentum representation with and without Dirac notation, and Fourier transform for relating the position space wavefunction with the momentum space wavefunction.

| State vector in Dirac notation | Generic quantum state vector: $|\Psi\rangle$ |
|---|---|
| Position representation | $\langle x|\Psi\rangle = \Psi(x)$ |
| Momentum representation | $\langle p|\Psi\rangle = \Phi(p)$ |
| State vector in Dirac notation | Position eigenstate with eigenvalue $x'$: $|x'\rangle$ |
| Position representation | $\langle x|x'\rangle = \delta(x-x')$ |
| Momentum representation * | $\langle p|x'\rangle = e^{-ipx'/\hbar}$ |
| State vector in Dirac notation | Momentum eigenstate with eigenvalue $p'$: $|p'\rangle$ |
| Position representation * | $\langle x|p'\rangle = e^{ip'x/\hbar}$ |
| Momentum representation | $\langle p|p'\rangle = \delta(p-p')$ |
| Fourier transform to relate momentum space wavefunction to position space wavefunction | $\Phi(p) = \langle p|\Psi\rangle = \int_{-\infty}^{\infty} \langle p|x\rangle \langle x|\Psi\rangle dx = \int_{-\infty}^{\infty} e^{-ipx'/\hbar} \Psi(x) dx$ |

* We note that for $\langle x|p'\rangle$ and $\langle p|x'\rangle$, some textbooks use certain pre-factors for "normalization", e.g., $\langle x|p'\rangle = \frac{1}{\sqrt{2\pi\hbar}} e^{ip'x/\hbar}$ (though, these states cannot be normalized). In the data presented here, student answers were counted as correct to these questions whether or not they included the pre-factors as long as they had the correct expression (e.g., $e^{ip'x/\hbar}$ for $\langle x|p'\rangle$).

### C. Methods used to identify student difficulties

To identify student difficulties, we administered open-ended and multiple-choice questions on quizzes and exams to upper-level undergraduate and Ph.D. students after traditional lecture-based instruction in relevant concepts and conducted individual interviews with students who were currently enrolled in a quantum mechanics course [79]. Additionally, during the development and validation of the QuILT as well as the corresponding pre-/post-test, semi-structured think-aloud interviews were conducted with 23 undergraduate students. During the interviews, the students worked through a particular version of the QuILT while verbalizing their thought process. They were not interrupted except at times when they became quiet for roughly 30 seconds in which case, they were prompted to verbalize what they were thinking about. Occasionally, if students' reasoning was not clear when they worked on a particular question, after they were finished, the interviewer asked the student to clarify or elaborate further so their though process was clear. These interviews provided valuable insights into common student difficulties (what knowledge resources get activated in response to different questions) as well as the associated reasoning patterns. We note that students who participated in these interviews were not a subset of the ones from which we collected quantitative data (Q1-Q7, see Section III.) The interviews were used to understand common student difficulties with these concepts as well as to make refinements to the QuILT based on student feedback from working through preliminary versions of the QuILT.

### D. Methods used for developing and validating the QuILT

Identifying common student difficulties as well as the associated reasoning patterns were invaluable for designing a QuILT that would be effective in helping students learn these concepts related to quantum state vectors in Dirac notation and as wavefunctions in the position and momentum representations. The QuILT uses a guided inquiry-based approach and consists of teaching-learning sequences in which different concepts build on each other to help students organize, extend, and repair their knowledge structure. Since the focus is on representations of quantum states and facility in translating between different representations, the QuILT includes many questions requiring students to translate quantum states from one representation to another and provides guidance and scaffolding support such that the tasks are within students' zone of proximal development. The QuILT explicitly incorporates many of the common difficulties, e.g., in the form of hypothetical student discussions in which one student is voicing a common incorrect type of reasoning for a specific question and another is voicing correct reasoning. Students are asked to identify who they agree with and why.

The development and validation of the QuILT and the corresponding pre-/post-test went through a cyclic, iterative process which included the following stages before the in-class implementation:

1. Development of a preliminary version based on a cognitive task analysis of the underlying knowledge from expert and student perspective (e.g., by researching student difficulties with relevant concepts, i.e., what knowledge resources get activated in response to different questions)
2. Implementation and evaluation of the different versions of the QuILT (along with the pre-test and post-test) by administering it individually to students (in one-on-one interviews) and obtaining feedback from faculty members who are experts in these topics and graduate students who conduct research on student understanding of quantum mechanics
3. Determining the impact of the QuILT on student learning in individual interviews and assessing what difficulties were not adequately addressed by the QuILT based upon the feedback obtained
4. Making refinements and modifications based on each feedback from the implementation and evaluation

### E. In-class implementation and evaluation of the QuILT

After we determined that the QuILT was effective in individual student interviews, it was administered to upper-level undergraduate and Ph.D. students. For the undergraduate students in the upper-level mandatory quantum physics course, a pretest on the topics was administered in class after traditional lecture-based instruction in relevant concepts and all students were given enough time to work through pretest. Students then worked through the QuILT in class in small groups and were given one week to finish the QuILT as a homework assignment. The pretest and QuILT counted as a part of their homework grade. The pretest was administered in class and was not returned to students. The posttest was then administered to the undergraduate students in class after students had turned in the homework and students were given enough time to work through it. The posttest was graded for correctness as a quiz. In addition, since the tutorial was part of the course material, the upper-level undergraduate students were told that topics in the QuILT could be on future exams.

For the first-year physics Ph.D. students, in the first two years of administration, the QuILT was used as part of a mandatory course for training teaching assistants in their first year (first semester) of the Ph.D. program. This course meets for one 2-hour class each week. In this course, the first-year Ph.D. students learned about various instructional strategies (including tutorial-based approaches to learning via engaging with the QuILT). They first worked on the pretest (all students were given enough time to work through the pretest), after which they worked on the QuILT in small groups in class to learn about the tutorial approach. They were given one week to work through the rest of the QuILT as a homework assignment. Then, the Ph.D. students were administered the posttest in class and they were given enough time to work through it. The Ph.D. students were given credit for completing the pretest, QuILT, and posttest, but they were not given credit for correctness. In other words, the Ph.D. students' scores on the posttest did not contribute to the final grade for the teaching assistant training class (which was a Pass/Fail course based upon completing activities and homework). In the last two years of administration, the Ph.D. students worked on the QuILT as part of a graduate level quantum mechanics course, and the administration was identical to that of the undergraduate students. They were given the pretest after traditional lecture-based instruction, then they worked on the QuILT in class in small groups and were given a week to complete the QuILT as a homework assignment. The posttest was then administered during the next week as a quiz after students had submitted their homework.

### III. STUDENT DIFFICULTIES

Before and throughout the development and validation of the QuILT and the corresponding pretest and posttest, we investigated student difficulties with representations of quantum states [79]. Table II shows the validated pre- and posttest questions related to representations of quantum states that were given to the students in in-class administration over a period of four years. Additionally, a multiple-choice question related to representation of quantum states was administered as part of a conceptual survey, called the Quantum Mechanics Formalism and Postulates Survey (QMFPS) [73]. We refer to this as the "retention question" because the QMFPS survey was administered during the last week of the semester. Thus, student performance on this question provides an indication of the extent to which students retained some of what they had learned about representations of quantum states. A total of 124 students who worked on the QuILT, 58 undergraduate students and 66 graduate students, also answered this retention question from the QMFPS survey at the end of the semester. Additionally, the same retention question was administered to undergraduate students from four separate large research universities (184 students who did not work on the QuILT) at the end of the semester. The difference in performance on the retention question for the QuILT and non-QuILT groups can provide some indication as to the effectiveness of the QuILT.

**Retention question** (correct answer shown in bold)**:** Choose all of the following statements that are correct about the position space and momentum space wavefunctions for this quantum state.
  (1) The position space wavefunction is $\Psi(x) = \langle x|\Psi\rangle$ where $x$ is a continuous index.
  (2) The momentum space wavefunction is $\Phi(p) = \langle p|\Psi\rangle$ where $p$ is a continuous index.
  (3) The momentum space wavefunction is $\Phi(p) = \int dx(-i\hbar\frac{\partial}{\partial x}\Psi(x))$
A. all of the above   B. 2 only   **C. 1 and 2 only**   D. 3 only   E. 1 and 3 only

**Table II**. Pre- and posttest questions involving quantum states that were administered to undergraduate students (UG) and Ph.D. students (G) and the number of students ($N$) answering the questions. We note that the instructions for students were that for Q4-Q7 written in Dirac notation, they should write it without using Dirac notation.

| Question | N |
|---|---|
| **Q1.** How would you obtain the wavefunction in position representation from $|\Psi\rangle$? | 85 UG<br>97 G |
| **Q2.** Write a momentum eigenstate with eigenvalue $p'$ in position representation.[*] | 85 UG<br>97 G |
| **Q3.** Write a momentum eigenstate with eigenvalue $p'$ in momentum representation. | 62 UG<br>68 G |
| **Q4.** $\langle x|x'\rangle = ?$ | 85 UG<br>97 G |
| **Q5.** $\langle p|p'\rangle = ?$ | 85 UG<br>97 G |
| **Q6.** $\langle x|p'\rangle = ?$ | 85 UG<br>97 G |
| **Q7.** $\langle p|x'\rangle = ?$ [1] | 85 UG<br>97 G |

[*] In one of the four years, pretest and posttest questions were slightly different, i.e., if students were asked to write a momentum eigenstate with eigenvalue $p'$ in position representation (i.e., $\langle x|p'\rangle$) in the pretest, the post-test asked them to write a position eigenstate with eigenvalue $x'$ in momentum representation (i.e., $\langle p|x'\rangle$). .

Below, we describe the difficulties found in written responses and interviews. Since students were significantly more likely to have difficulties after traditional lecture-based instruction (pretest) than after engaging with the QuILT (posttest), the pretest results are shown in Table III. Table III also shows results for the retention question.

**Table III**. Percentages of undergraduate students (UG) and Ph.D. students (G) who correctly answered questions Q1-Q7 and percentage of undergraduate students who selected each answer choice in the retention question at the end of the semester (these students did not work on the QuILT).

| Question | Percent correct (Q1-Q7) or percent of students who selected each answer choice for the retention question |
|---|---|
| Q1 | 43% UG, 77% G |
| Q2 | 10% UG, 55% G |
| Q3 | 11% UG, 54% G |
| Q4 | 26% UG, 81% G |
| Q5 | 25% UG, 79% G |
| Q6 | 13% UG, 65% G |
| Q7 | 13% UG, 62% G |
| Retention question | A (37%), B (2%), **C (38%)**, D (10%), E (14%) |

### A. Undergraduate student difficulties

The difficulties reported here are consistent with those reported in a conference proceedings paper [79] but are for a larger number of students.

**Discrepancy in student performance on questions about a generic state vector $|\Psi\rangle$ posed in different formats:** Table III shows that, on the retention question posed in a multiple-choice format, 89% of the undergraduate students recognized that a generic state vector $|\Psi\rangle$ written in the position representation is $\Psi(x) = \langle x|\Psi\rangle$ based on selecting options A, C, or E that included statement (1). We note that the guessing probability for selecting statement (1) is 60% because statement (1) was included in three answer options. The same is true for statement (2). The percentage of 89% is larger than expected from random guessing on the retention question. However, on Q1 (see Table III), when students are asked to write a state vector $|\Psi\rangle$ in position representation in a free-response format, student performance is significantly lower. Table III shows that on Q1, only 43% of the undergraduate students provided a correct answer. We note that answers were considered correct if students used Dirac notation, e.g., $\langle x|\Psi\rangle$, or if they used algebraic notation, e.g., $\Psi(x)$, or if they stated that the generic state $|\Psi\rangle$ must be projected onto the position basis. The dichotomy in student performance on Q1 vs. the retention question may be due to the problem statement which provides some cues as well as the multiple-choice format of the retention question since students do not have to generate a response. In particular, the retention question includes the correct statement that the position space wavefunction is $\Psi(x) = \langle x|\Psi\rangle$, and students are only required to recognize that this statement is correct. However, Q1 provides no scaffolding and asks students to generate an expression for the wavefunction in the position representation from $|\Psi\rangle$. The difference in student performance in Q1 vs. the retention question indicates that student performance is dependent on whether they must generate or only recognize an answer: they are able to recognize the correct expression, but much less able to generate it. This suggests that they are currently developing expertise. We also note that the same question is more challenging in the context of momentum: as shown in Table III, 77% of the students recognized that a generic state vector $|\Psi\rangle$ in momentum representation is $\Phi(p) = \langle p|\Psi\rangle$ based on selecting answer option A, B, or C that included statement (2).

Below, we discuss some difficulties that undergraduate students had with the questions in Table II:

**Incorrectly claiming that writing a generic state vector $|\Psi\rangle$ in position or momentum representation involves the position or momentum operator, respectively:** As noted, many undergraduate students had difficulty on Q1 that asked how one would obtain the wavefunction in position representation from $|\Psi\rangle$. Many students included the position operator in their response (18% of students as shown in Table IV). Below, we provide a few examples:

- $\hat{x}|\Psi\rangle = x|\Psi\rangle$,
- $\hat{x}|\Psi\rangle = \langle x|\Psi\rangle$
- $\langle x|\Psi\rangle = \int \hat{x}^*\Psi dx = \int x\,\Psi dx$
- $\hat{x}\,\Psi(x) = \langle x|\hat{x}|\Psi\rangle$

We found similar difficulties when students attempted to write a generic state $|\Psi\rangle$ in momentum representation, namely, that students included the momentum operator in their response. For example: $\langle p|\Psi\rangle = \int \hat{p}^*\Psi dx = \int i\hbar\,\partial/\partial x\,\Psi dx$. In interviews, students sometimes incorrectly claimed that writing a generic state $|\Psi\rangle$ in the position representation meant that one must act with the position operator $\hat{x}$ on that state $|\Psi\rangle$. Furthermore, even among students who noted correctly that $\langle x|\Psi\rangle = \Psi(x)$ and that $\Psi(x)$ is the wavefunction in the position representation, when they were asked to describe what $\langle x|\Psi\rangle$ physically means, some wrote similar statements as those included earlier that involved the position operator. For example: "$\langle x|\Psi\rangle$ is just $\int x^*\Psi dx = \Psi$ in position basis" or "$\langle x|\Psi\rangle$ is the measurement of $|\Psi\rangle$ in position, it yields a position eigenstate of the system at the time of measurement." Similarly, when asked what $\langle p|\Psi\rangle$ physically means, another student wrote: "$\langle p|\Psi\rangle = \int \Psi\left(-\frac{\hbar}{i}\frac{\partial}{\partial x}\right)\Psi dx = -\frac{\hbar}{i}\int \Psi\left(\frac{\partial}{\partial x}\right)\Psi$." In other words, students often confused projecting a state vector $|\Psi\rangle$ along an eigenstate of position (or momentum) with acting with the position (or momentum) operator on the state vector $|\Psi\rangle$.

In the retention question, this type of difficulty is explicitly brought out because students are asked to evaluate the correctness of the statement "The momentum space wavefunction is $\Phi(p) = \int dx(-i\hbar\frac{\partial}{\partial x}\Psi(x))$". Table IV shows that 61% of the students selected an answer A, D, or E that included this statement. The statement is incorrect because in order to obtain $\Phi(p)$ from $\Psi(x)$, a Fourier transform must be used, i.e., $\Phi(p) = \langle p|\Psi\rangle = \int dx\langle p|x\rangle\langle x|\Psi\rangle = \int dx\,e^{-ipx/\hbar}\Psi(x)$.

**Incorrectly claiming that writing a position or momentum eigenstate in position or momentum representation involves the position or momentum operator:** Similar to the difficulty in which students' expressions for a generic state $|\Psi\rangle$ in the position representation involved the position operator, students' expressions for the momentum eigenstates in the position or momentum representation often involved the momentum operator. On Q2, Table IV shows that 40% of the undergraduate students claimed, e.g., that writing a momentum eigenstate with eigenvalue $p'$ in the position representation involves the momentum operator. Examples include: $\hat{p}|\Psi\rangle = p'|\Psi\rangle$, $\hat{p}|\Psi\rangle_x$, $\hat{p}|x\rangle = p'e^{-ip'x/\hbar}$, $\langle \Psi|\hat{p}|x'\rangle$, $\langle x|\hat{p}|p'\rangle$, $|p\rangle = i\hbar p'|x\rangle$, $i\hbar\frac{\partial}{\partial x}|\Psi\rangle$,

$i\hbar \frac{\partial}{\partial x'}$. Similarly, on Q3, Table IV shows that 50% of the undergraduate students claimed that the momentum operator is included in the expression for a momentum eigenstate with eigenvalue $p'$ in the momentum representation. Examples include: $\hat{p}|\Psi\rangle = p'|\Psi\rangle$, $\hat{p}|\Psi\rangle = p'|p'\rangle$, and $\hat{p}|p'\rangle = p'|p'\rangle$. Interviews suggest that at least some students thought that this question relates to a measurement of momentum ($\hat{p}|\Psi\rangle = p'|\Psi\rangle$), or that this also collapses the state to a momentum eigenstate ($\hat{p}|\Psi\rangle = p'|p'\rangle$). Others simply confused writing a momentum eigenstate with eigenvalue $p'$ in the momentum representation with the eigenvalue equation for the momentum operator ($\hat{p}|p'\rangle = p'|p'\rangle$). Similar difficulties have been reported in the context of quantum measurement. For example, students sometimes incorrectly claim that the Hamiltonian operator $\hat{H}$ acting on a generic state $|\Psi\rangle$ corresponds to the measurement of energy and should yield $\hat{H}|\Psi\rangle = E|\Psi\rangle$ or $\hat{H}|\Psi\rangle = E_n|\Psi_n\rangle$ [20, 35, 71, 111], where $E$ or $E_n$ is the energy measured and $|\Psi_n\rangle$ is the energy eigenstate in which the state of the quantum system collapsed. While the measurement of a physical observable corresponds to the quantum state collapsing into one of the many eigenstates of the operator corresponding to the observable measured (e.g., $\hat{H}$ corresponds to the observable energy), the measurement process cannot be described by the operator corresponding to a physical observable acting on the state before measurement. We have argued elsewhere that these kinds of difficulties are related to a reasoning primitive that students use, namely, "operator measures observable", i.e., acting with an operator corresponding to an observable on a state corresponds to a measurement [112].

**Table IV.** Undergraduate student difficulties with writing quantum states using position and momentum representation (these students had not worked on the QuILT). We note that the percentages in this table are slightly different from the ones in Ref. [79] because the data presented here includes roughly twice as many students as the one in Ref. [79].

| Difficulty | Question | % |
|---|---|---|
| Claiming that the expression for a generic state vector $|\Psi\rangle$ in position or momentum representation involves the position or momentum operator, respectively | Retention question choosing option (3) | 61 |
| | Q1 | 18 |
| Claiming, e.g., that the expression for a momentum eigenstate with eigenvalue $p'$ in the position representation involves the momentum operator | Q2 | 40 |
| Claiming that the expression for a momentum eigenstate with eigenvalue $p'$ in the momentum representation involves the momentum operator | Q3 | 50 |
| Confusing a state with an operator resulting, e.g., in $\langle x|x'\rangle = x'$ or $x'$ multiplied by either a state or a function involving $x$ | Q4 | 19 |
| Confusing a state with an operator resulting, e.g., in $\langle p|p'\rangle = p'$ or $p'$ multiplied by either a state or a function involving $p$ | Q5 | 13 |
| Claiming that $\langle x|x'\rangle = 1$ or 0 | Q4 | 11 |
| Claiming that $\langle p|p'\rangle = 1$ or 0 | Q5 | 10 |
| Claiming that $\langle x|p'\rangle = 0$ | Q6 | 10 |
| Claiming that $\langle p|x'\rangle = 0$ | Q7 | 8 |

Q4 asked students to write an expression for $\langle x|x'\rangle$ without using Dirac notation. In response to Q4, 19% of the students had difficulty distinguishing between a state and an operator and wrote $\langle x|x'\rangle = x'$ or $x'$ multiplied by either a state or a function involving $x$. Interviews confirmed this confusion as students sometimes took the bra state $\langle x|$ to be the same as the position operator $\hat{x}$, so, e.g., $\langle x|x'\rangle$ became $\hat{x}|x'\rangle$. They then acted the momentum operator on $|x'\rangle$ and got, e.g., either $x'|x'\rangle$, or just $x'$. Students exhibited a similar confusion in Q5 between the bra state $\langle p|$ and the momentum operator $\hat{p}$. For example: $\langle p|p'\rangle = \hat{p}|p'\rangle = p'$, and sometimes the end result was $p'|p'\rangle$ or $p'$ multiplied by a function involving $p$.

**Assuming $\langle x|x'\rangle = 1$ or 0 (or $\langle p|p'\rangle = 1$ or 0):** Table III shows that, around one quarter of the undergraduate students answered Q4 and Q5 correctly. Interviews suggest that students sometimes incorrectly invoked orthogonality, for example, claiming that $\langle x|x'\rangle = 1$ or 0 (11% of students) or that that $\langle p|p'\rangle = 1$ or 0 (10% of students). These students thought that unless $x = x'$ or $p = p'$, the inner products would be zero. In other words, some students confused the Kronecker delta with the Dirac delta function. In these cases, since momentum and position have continuous eigenvalues, one should write the inner products as Dirac delta functions, whereas the Kronecker delta is for inner products of eigenstates that have discrete eigenvalues.

**Assuming $\langle x|p'\rangle = 0$ or $\langle p|x'\rangle = 0$:** On Q6 and Q7, Table III shows that only 13% of the students correctly recalled that $\langle x|p'\rangle = e^{ip'x/\hbar}$ and $\langle p|x'\rangle = e^{-ipx'/\hbar}$. (We note that students were given full credit if they included a constant pre-factor for normalization purposes or if they had an incorrect sign in the exponential.) Similar to $\langle x|x'\rangle$ and $\langle p|p'\rangle$, students sometimes invoked orthogonality. On Q6 for example, 10% of the students wrote that $\langle x|p'\rangle = 0$ and on Q7, 8% of the students wrote that $\langle p|x'\rangle = 0$. Interviews confirmed this as some students claimed that eigenstates of position and momentum are orthogonal. Other students claimed that the eigenstates are incompatible (somewhat similar idea but it is also reminiscent of operators being compatible or incompatible) and thought that the inner products of one with the other did not make sense. But because they were asked what those inner products were, they thought that they can only be zero because position and momentum have "nothing to do with each other". Prior research has identified similar difficulties when investigating student understanding of spin-1/2 systems, where students for example claim that eigenstates of $\hat{S}_x$ and $\hat{S}_z$ are orthogonal [38].

### B. Ph.D. student difficulties

**Discrepancy in Ph.D. students' performance on questions about a generic state vector $|\Psi\rangle$ posed in different formats:** Table III shows that, on Q1-Q7 on the pretest, Ph.D. students had significantly better average performance on the pretest compared to the undergraduate students. For example, on Q6, 65% of the Ph.D. students correctly wrote that $\langle x|p'\rangle = e^{ip'x/\hbar}$ (as compared to only 13% of the undergraduate students). But only 55% correctly answered Q2 (write a momentum eigenstate with eigenvalue $p'$ in the position representation). In other words, 10% of the Ph.D. students knew that $\langle x|p'\rangle = e^{ip'x/\hbar}$, but did not know that $\langle x|p'\rangle$ refers to a momentum eigenstate with eigenvalue $p'$ in position representation. Similarly, on Q5, 79% of Ph.D. students correctly wrote that $\langle p|p'\rangle = \delta(p-p')$, which is significantly higher than the undergraduate student performance on the same question (25%). But only 54% correctly answered Q3 (write a momentum eigenstate with eigenvalue $p'$ in momentum representation). In other words, 25% of the Ph.D. students knew that $\langle p|p'\rangle = \delta(p-p')$ but did not know that $\langle p|p'\rangle$ refers to a momentum eigenstate with eigenvalue $p'$ in momentum representation. On Q2 and Q3, the Ph.D. students had similar difficulties as the undergraduate students as discussed earlier.

This dichotomy in Ph.D. students' performance between Q2-Q3 and Q5-Q6 suggests that while the majority of Ph.D. students are able to recall how to convert certain expressions from Dirac notation to algebraic notation, many are not aware of the physical meaning of those expressions and struggle to generate them. Writing a momentum eigenstate with eigenvalue $p'$ in the position (or momentum) representation requires one to understand the various symbols involved, both in Dirac notation as well as in position (or momentum representation). If instruction in a graduate level QM course primarily focuses on solving problems that require merely recalling these types of expressions reproducing them on homework and exams (rather than also requiring understanding what the expressions actually mean), it is unlikely that students can develop a functional understanding of these expressions.

We note that the data presented here comes from students at a large public research university who generally perform quite well in their quantum mechanics courses. And even these students show persistent difficulties with translating between different representations of state vectors, as well as recognizing conceptually the meaning of various related expressions in Dirac notation, e.g., $\langle x|x'\rangle$ refers to a position eigenstate with eigenvalue $x'$ in position representation. This suggests that (1) students at other less selective institutions are likely to struggle even more with these issues, and (2) there is a need to incorporate evidence-based instructional materials to help students learn these concepts in different representations.

### IV. ADDRESSING STUDENT DIFFICULTIES VIA THE QUILT

The entire Dirac notation QuILT can be found on PhysPORT [113]. Here we summarize and illustrate with examples how the QuILT is designed to address some of the difficulties discussed in the preceding section with quantum states.

#### A. Addressing student difficulties with writing state vectors in Dirac notation and as wavefunctions in position and momentum representations via the QuILT

**Addressing the difficulties with writing a generic state vector $|\Psi\rangle$ in the position or momentum representation.** As noted earlier, students sometimes involved the position operator or momentum operator in their expressions for writing state vector $|\Psi\rangle$ in the position or momentum representation. The following questions are part of a guided inquiry-based learning sequence in the QuILT that strives to help students learn how to write a generic quantum state vector $|\Psi\rangle$ in the position or momentum representation as an infinite dimensional column vector or as a wavefunction (we note that in the excerpts below, the text written in italics is taken from the QuILT and the text in parentheses that is not italicized provides correct answers; in the case of multiple-choice questions, the correct answer is in bold):

- *Choose all of the following statements that are correct about the generic state vector $|\Psi\rangle$*
  - (I) *$\Psi(x) = \langle x|\Psi\rangle$ is a representation of the state vector $|\Psi\rangle$ in the position representation and is called the position space wavefunction.*
  - (II) *$\Phi(p) = \langle p|\Psi\rangle$ is a representation of the state vector $|\Psi\rangle$ in the momentum representation and is called the momentum space wavefunction.*

(III) The state vector $|\Psi\rangle$ can be written as a column vector once a basis has been chosen.
(All statements are correct)
- Consider the following conversation between two students:
  - Student A: In the preceding question, we can also write $\Psi(x)$ as $|\Psi\rangle = \Psi(x)$.
  - Student B: I disagree. $|\Psi\rangle \doteq \Psi(x)$, where the $\doteq$ sign means that the equality is valid only with respect to a chosen basis. $\Psi(x)$ is a representation of $|\Psi\rangle$ in position representation when we choose a complete set of position eigenstates, $|x\rangle$, as our basis vectors, so we can write $\Psi(x) = \langle x|\Psi\rangle$.

With whom do you agree? Explain your reasoning.
(Student B is correct)

Students are provided additional scaffolding support through the following types of questions:

*Consider the graphs and statement made by Student A:*

*Student A: For a continuous variable like position, the column vector representation $|\Psi\rangle \doteq \begin{pmatrix} \langle x_1|\Psi\rangle \\ \langle x_2|\Psi\rangle \\ \vdots \end{pmatrix}$ is not convenient because we cannot write down an infinite number of components. We can translate from the column vector representation of discrete points (shown in the top figure-Fig. 1) to a continuous set of numbers which is called the quantum mechanical wavefunction (shown in the bottom figure-Fig. 2). The wavefunction is an infinite collection of numbers that represents the quantum state vector in terms of position eigenstates.*

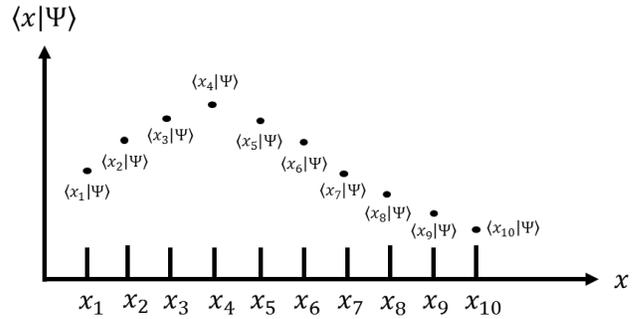

Fig.1. Visualization of $\Psi(x)$ for a discrete set of basis vectors

*Explain why you agree or disagree with Student A's statement.*

(Agree because even though the expansion of $|\Psi\rangle$ using position eigenstates is an integral instead of a sum, one can still envision this expansion as an infinite column vector in which the rows of the column vector correspond to the basis vectors $|x\rangle$.)

Later in the QuILT, another student discussion helps students recognize that for the column vector mentioned earlier, namely, $|\Psi\rangle \doteq \begin{pmatrix} \langle x_1|\Psi\rangle \\ \langle x_2|\Psi\rangle \\ \vdots \end{pmatrix} = \begin{pmatrix} \Psi(x_1) \\ \Psi(x_2) \\ \vdots \end{pmatrix}$ one can replace $x_1 = \Delta x$, $x_2 = 2\Delta x$, etc., and take the limit as $\Delta x \to 0$. In other words, $\Psi(x)$ can be thought of as a column vector with continuous position eigenvalues $x$.

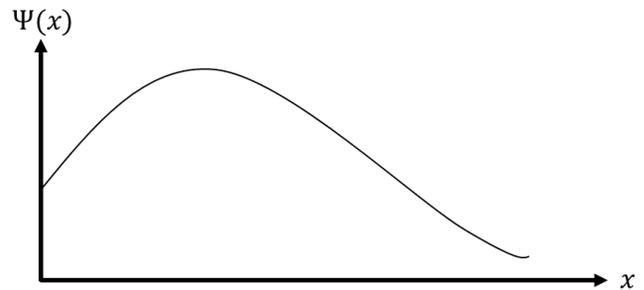

Fig.2. Visualization of $\Psi(x)$ for a continuous set of basis vectors

Following this guided inquiry-based learning sequence and peer discussion, additional questions that build on the preceding ones in the QuILT provide scaffolding to help students learn how to write a state vector $|\Psi\rangle$ in the position representation in Dirac notation or as a wavefunction in algebraic notation without Dirac notation. The scaffolding is gradually reduced and then students are asked to write on their own a state vector $|\Psi\rangle$ in the momentum representation as $\langle p|\Psi\rangle$, which is the wavefunction in momentum representation written in Dirac notation or as a wavefunction in algebraic notation.

**Difficulty writing a position eigenstate $|x'\rangle$ or momentum eigenstate $|p'\rangle$ in the position or momentum representation.** In the preceding section, we discussed that students had difficulty writing, e.g., momentum eigenstates with eigenvalue $p'$ in the position and momentum representations, which often involved students incorrectly using operators in their expressions. Similar difficulties were found in interviews in writing e.g., position eigenstates with eigenvalue $x'$ in the position and momentum representations. The following excerpt from a guided inquiry-based learning sequence in the QuILT strives to help students learn how to write the position operator acting on the position eigenstate in the position representation and differentiate between the concepts of the position operator and the state in the position representation:

- Consider the following conversation between Student A and Student B.
  - Student B: I don't understand how $\hat{x}|x'\rangle = x'|x'\rangle$ and $\langle x|\hat{x}|x'\rangle = x\delta(x - x') = x'\delta(x - x')$ can both be correct. They are two different eigenvalue equations for the same operator $\hat{x}$, so shouldn't one of them be incorrect?
  - Student A: Actually, $\hat{x}|x'\rangle = x'|x'\rangle$ and $x\delta(x - x') = x'\delta(x - x')$ convey the same information. The former is the eigenvalue equation for the position operator $\hat{x}$ in Dirac notation, and the latter is the eigenvalue equation for the position operator in the position representation. In the position representation, $\hat{x}$ is equivalent to a multiplication by $x$. We can also write $\langle x|\hat{x}|x'\rangle = x\delta(x - x') = x'\delta(x - x')$ since $\langle x|x'\rangle = \delta(x - x')$, which is zero for all position eigenvalues except when $x = x'$.

*With whom do you agree? Explain.*
(Agree with Student A)

Following this guided inquiry-based sequence and peer discussion, further scaffolding is provided to help students check their work and reconcile possible differences between their responses.

In the preceding section, we discussed that some students invoked an incorrect orthonormality condition and claimed that $\langle x|x'\rangle = 1$. Others invoked an incorrect orthogonality condition and claimed that, e.g., $\langle p|x'\rangle = 0$. The following guided inquiry-based sequence in the QuILT strives to help students translate between wavefunction in position representation (in Dirac notation and algebraic notation) and reason about the fact that a position eigenstate in position representation, $\langle x|x'\rangle$, is a Dirac delta function in algebraic notation and the position eigenstate in the momentum representation, $\langle p|x'\rangle$, is $\frac{e^{-ipx'/\hbar}}{\sqrt{2\pi\hbar}}$:

- *Which one of the following is the eigenstate of position $|x'\rangle$ with eigenvalue $x'$ written in position representation, i.e., $\langle x|x'\rangle$?*
  - (a) $\langle x|x'\rangle = \Psi(x')$
  - **(b) $\langle x|x'\rangle = \delta(x - x')$**
  - (c) $\langle x|x'\rangle = \frac{e^{ipx'/\hbar}}{\sqrt{2\pi\hbar}}$
  - (d) $\langle x|x'\rangle = \Psi(x)$
- *Consider the following conversation between two students:*
  - *Student 1: The position eigenfunction should always be a delta function whether we write it in position or momentum representation.*
  - *Student 2: I disagree. $\langle x|x'\rangle$ cannot be the same as $\langle p|x'\rangle$ because when position eigenstate $|x'\rangle$ with eigenvalue $x'$ is written by choosing position eigenstates $|x\rangle$ as basis vectors, we obtain $\langle x|x'\rangle = \delta(x - x')$ which is the position eigenfunction in the position representation. When the position eigenstate $|x'\rangle$ is written by choosing momentum eigenstates $|p\rangle$ as basis vectors, we obtain $\langle p|x'\rangle$, which is the position eigenfunction in the momentum representation. However, $\langle p|x'\rangle$ is not a delta function. The position eigenfunction is only a delta function in the position representation, but not in the momentum representation.*
  
  *With whom do you agree? Explain your reasoning.*
  (Student 2 is correct.)

After this question, students are asked a similar question about $\langle p|x\rangle$ written in algebraic notation. The options are $\frac{1}{\sqrt{2\pi\hbar}}e^{-ipx/\hbar}$, $\delta(p - x)$, $\delta(p - p')$, $\delta(x - x')$ (the first option is correct) and follow-up questions are used to help them figure out the correct answer and reconcile any differences. Students also verify that the expression they chose for the momentum eigenfunction in the position representation satisfies the eigenvalue equation for momentum operator in the position representation. The QuILT also provides scaffolding support to help students learn how to use a Fourier transformation to translate between the position and momentum representations.

As noted in the preceding section, some students were inconsistent in their responses to questions posed in different formats. For example, some students knew that $\langle p|p'\rangle = \delta(p - p')$, but when they were asked to write a momentum eigenstate with eigenvalue $p'$ in the momentum representation, they struggled to do so either using Dirac notation or algebraic notation. The following question from the QuILT asks students to consider these issues and then helps them associate a conceptual meaning to the expressions, e.g., in Dirac notation $\langle p|p'\rangle$.

- *Which one of the following is the eigenstate of momentum $|p'\rangle$ with eigenvalue $p'$ in momentum representation, i.e., $\langle p|p'\rangle$ (momentum eigenfunction in momentum representation)?*
  - (a) $\langle p|p'\rangle = \Phi(p')$
  - **(b) $\langle p|p'\rangle = \delta(p - p')$**
  - (c) $\langle p|p'\rangle = \frac{e^{ip'x/\hbar}}{\sqrt{2\pi\hbar}}$
  - (d) $\langle p|p'\rangle = \frac{e^{-ip'x/\hbar}}{\sqrt{2\pi\hbar}}$

Lastly, two more student discussions are used to help solidify students' understanding of all the different inner products between position and momentum eigenstates. (We note that in the excerpts below, some symbols which were defined in the QuILT but not in this article are removed.)

*Consider the following conversation between two students about a position eigenfunction and a momentum eigenfunction in the position representation:*

- Student 1: A position eigenstate is represented in position representation as $\langle x|x'\rangle = \delta(x-x')$ such that $\delta(x-x')$ is a special type of position space wavefunction $\Psi(x) = \langle x|\Psi\rangle$ in which position has a definite value of $x'$.

- Student 2: I agree with you. In addition, a momentum eigenstate is represented in position representation as $\langle x|p'\rangle = \frac{e^{\frac{ip'x}{\hbar}}}{\sqrt{2\pi\hbar}}$. $\frac{e^{\frac{ip'x}{\hbar}}}{\sqrt{2\pi\hbar}}$ is another special type of position space wavefunction $\Psi(x) = \langle x|\Psi\rangle$ where momentum has a definite value $p'$. Instead of being sharply peaked like a delta function, a momentum eigenfunction in the position representation is spread out as $\frac{e^{\frac{ip'x}{\hbar}}}{\sqrt{2\pi\hbar}}$ (which is a linear combination of sine and cosine functions over all position with a definite momentum $p'$ and wave number $k' = \frac{p'}{\hbar}$) and the probability density is uniform.

*Do you agree with Student 1, Student 2, both, or neither? Explain your reasoning.*
(Agree with both)

Another analogous discussion focuses on momentum representation as follows:
*Consider the following conversation between two students:*

- Student 1: $\langle p|p'\rangle = \delta(p-p')$ is a special type of momentum space wavefunction $\Phi(p) = \langle p|\Psi\rangle$, in which momentum has a definite value $p'$. A momentum eigenfunction in the momentum representation, $\delta(p-p')$, is a very sharply peaked wavefunction about $p = p'$ in momentum representation.

- Student 2: I agree with you. In addition, a position eigenstate $|x'\rangle$ with eigenvalue $x'$ represented in momentum representation is $\langle p|x'\rangle = \frac{e^{\frac{-ipx'}{\hbar}}}{\sqrt{2\pi\hbar}}$. $\frac{e^{\frac{-ipx'}{\hbar}}}{\sqrt{2\pi\hbar}}$ is also a special type of momentum space wavefunction $\Phi(p) = \langle p|\Psi\rangle$ in which position has a definite value $x'$ ($x'$ is fixed in $\frac{e^{\frac{-ipx'}{\hbar}}}{\sqrt{2\pi\hbar}}$, which is a function of $p$).

*Do you agree with Student 1, Student 2, neither, or both? Explain your reasoning.*
(Agree with both)

## V. IN-CLASS EVALUATION OF THE QUILT

After we determined that the QuILT was effective in individual student interviews, it was administered to upper-level undergraduate students and Ph.D. students along with the validated pretest and posttest to evaluate the effectiveness of the QuILT in the in-class environment The pretest was administered to students after they had traditional instruction in relevant concepts, and the posttest was administered after they worked on the QuILT. The results are shown in Table V. The number of students on the posttest does not match the pretest because students' scores on the posttest were not counted if they did not work through the entire tutorial and some of the questions were not administered every year. Average normalized gain [114] is commonly used to measure how much the students learned and takes into account their initial scores on the pretest. It is defined as $\langle g\rangle = (\%\langle post\rangle - \%\langle pre\rangle)/(100 - \%\langle pre\rangle)$, in which $\langle post\rangle$ and $\langle pre\rangle$ are the final (post) and initial (pre) class averages, respectively. The average normalized gain from the pretest to the posttest was 0.75 for undergraduate students and 0.76 for Ph.D. students, which indicate significant learning from the QuILT. We also calculated the effect size denoted by $d$ in the form of Cohen's $d$ ($d = (\mu_1 - \mu_2)/\sigma_{pooled}$, where $\mu_1$ and $\mu_2$ are the averages of the two groups being compared and $\sigma_{pooled} = \sqrt{(\sigma_1^2 + \sigma_2^2)/2}$, where $\sigma_1$ and $\sigma_2$ are the standard deviations of the two groups) [115]. The average effect size on questions Q1-Q7 is 1.6 for undergraduate students (which is considered a large effect size) and 0.60 for Ph.D. students (which is considered a moderate effect size) [115]. The effect size is larger for undergraduates compared to graduate students primarily because the undergraduates had a significantly lower pretest score.

**Table V.** Average pretest and posttest scores and normalized gain <g> of undergraduate (UG) students and Ph.D. (G) students on posttest questions Q1-Q7.

|    | **Pretest Average Score (total number of students)** | **Posttest Average score (total number of students)** | **<g>** | **Effect size *d*** |
|----|------|------|------|------|
| Q1 | 40% UG (85), 79% G (97) | 79% UG (58), 91% G (66) | 0.65 UG, 0.57 G | 0.87 UG, 0.36 G |
| Q2 | 14% UG (85), 58% G (97) | 80% UG (58), 92% G (66) | 0.76 UG, 0.81 G | 1.94 UG, 0.48 G |
| Q3 | 14% UG (62), 57% G (68) | 68% UG (82), 76% G (95) | 0.63 UG, 0.44 G | 1.38 UG, 0.42 G |
| Q4 | 35% UG (85), 81% G (97) | 87% UG (82), 98% G (95) | 0.80 UG, 0.89 G | 1.37 UG, 0.58 G |
| Q5 | 29% UG (85), 79% G (97) | 89% UG (82), 97% G (95) | 0.85 UG, 0.86 G | 1.62 UG, 0.57 G |
| Q6 | 15% UG (85), 68% G (97) | 83% UG (82), 93% G (95) | 0.80 UG, 0.78 G | 1.89 UG, 0.75 G |
| Q7 | 15% UG (85), 64% G (97) | 82% UG (82), 98% G (95) | 0.79 UG, 0.94 G | 1.86 UG, 0.98 G |

Additionally, as mentioned earlier, at the end of the term, students were administered a conceptual QMFPS survey on formalism and postulates of quantum mechanics. One of the questions on the QMFPS is the retention question, and it provides some information as to the extent to which students retained some of the learning from the QuILT. Additionally, since this question was administered to both students who worked on the QuILT and students who did not (from four different large research universities) at the end of the term, this comparison is valuable to gauge the effectiveness of the QuILT. The results shown in Table VI indicate that students who worked on the QuILT performed significantly better than students who did not.

**Table VI.** Distribution of student responses on the retention question from the QMFPS survey administered at the end of the semester for students who did not work through the QuILT (non-QuILT group from four different large research public universities) and students who worked through the QuILT (QuILT group). Note that in the QuILT group, there were 58 undergraduate students and 66 graduate students who were combined due to their comparable performance.

| Question | **Non-QuILT Group** (N = 184 UG) | **QuILT Group** (N = 124 = 58 UG + 66 G) |
|---|---|---|
| **Retention Question** | A (37%), B (2%), **C (38%)**, D (10%), E (14%) | A (20%), B (0%), **C (75%)**, D (2%), E (2%) |

## VI. SUMMARY

In quantum mechanics, multiple types of representations are used to reason about quantum concepts. Learning Dirac notation and how to translate a state vector written in Dirac notation to wavefunctions using position and momentum representation and vice versa, is an important step in progressing from a novice to an expert in quantum mechanics. In this study we found that upper-level undergraduate and graduate student responses in quantum mechanics indicate they had difficulties related to translating quantum state vectors in Dirac notation and wavefunction in the position and momentum representations (in Dirac notation and algebraic notation) after traditional lecture-based instruction. Student responses were also dependent on whether the questions were asking them to generate an expression or recognize an expression provided as they struggled to translate between different representations. In particular, after traditional lecture-based instruction in relevant concepts, many students had difficulties with writing quantum states in the position and momentum representations if the question was asking them to generate the expression without any hint (e.g., an expression written in Dirac notation that needs to be converted to algebraic notation without Dirac notation). To help students develop expertise and facility with these representations of quantum states, the QuILT provides students multiple opportunities to answer questions focused on translating between representations using guided inquiry-based teaching learning sequences in which different questions build on each other. This gives students practice with building facility with a number of semiotic resources [103] necessary for a holistic understanding of these quantum concepts. Each representation gives students a different 'view' of the concept thus helping them develop a functional understanding of these concepts. Additionally, since the QuILT was developed using research on what students were able to do after traditional lecture-based instruction and what they struggled with, the guided inquiry-based sequences that provide scaffolding strive to remain in students' ZPD. The findings of this investigation are encouraging in that both upper-level undergraduate students and graduate students who engaged with the research-based QuILT were able to better translate between various representations, e.g., Dirac notation and wavefunction in position and momentum representations.

Finally, we note that the entire tutorial is focused on infinite-dimensional Hilbert spaces and student proficiency in translating between different representations in such spaces. Thus, it is designed to help students build proficiency beyond two-level systems involving infinite-dimensional Hilbert space, which will continue to be very important for physicists involved in the second quantum revolution.

## ACKNOWLEDGEMENTS

We thank the students and faculty who helped with this study. We are also grateful to NSF (the National Science Foundation) for award PHY-2309260.